 \documentclass[12pt]{article}
 \usepackage[top=1in,bottom=1in,left=1.5in,right=1.5in]{geometry}
 \usepackage{amsmath,amssymb}
 \usepackage{latexsym}
 \usepackage[dvips]{pstricks} 
 \usepackage[dvips]{graphicx}

 \newcommand{\N}{\mathbb{N}}
 \newcommand{\R}{\mathbb{R}}
 
 \newcommand{\Z}{\mathbb{Z}}

 \newcommand{\bP}{\mathbf{P}}
 
 \renewcommand{\u}{\mathbf{u}}

 \newcommand{\U}{\mathbf{U}}

 \newcommand{\x}{\mathbf{x}}
 \newcommand{\X}{\mathbf{X}}
 \newcommand{\y}{\mathbf{y}}
 
 \newcommand{\1}{\mathbf{1}}
 \newcommand{\0}{\mathbf{0}}

 \newcommand{\cC}{\mathcal{C}}

 \newcommand{\cP}{\mathcal{P}}

 \newcommand{\rB}{\mathrm{B}}
 
 \newcommand{\rM}{\mathrm{M}}

 \newcommand{\rS}{\mathrm{S}}

 \newcommand{\hs}{\hspace*{\parindent}}
 \newcommand{\proof}{\hs \textbf{Proof.\ }}
 \newcommand{\outproof}{\hs \textbf{Outline of Proof.\ }}
 \newcommand{\tr}{\mathop{\mathrm{tr}}\nolimits}

 \newcommand{\trans}{^\top}
 \newcommand{\qed}{\hspace*{\fill} $\Box$\\}

 \newcommand{\vol}{\mathrm{vol}}
 \newtheorem{theo}{\bfseries \hs Theorem}[section]

 \newtheorem{lemma}[theo]{\bfseries \hs Lemma}

 \numberwithin{equation}{section}

\begin{document}

 \title{Graph isomorphism and \\ volumes of convex bodies}

 \author{
 Shmuel Friedland\\
 Department of Mathematics, Statistics and Computer Science\\
 University of Illinois at Chicago\\ Chicago, Illinois 60607-7045,
 USA\\ \texttt{E-mail: friedlan@uic.edu}
 }

 \date{March 13, 2009}

 \maketitle

 \begin{abstract}
  We show that a nontrivial graph isomorphism problem of two undirected graphs,
  and more generally, the permutation similarity of two given $n\times n$
  matrices, is equivalent to equalities of volumes of the induced three convex
  bounded polytopes intersected with a given sequence of balls,
  centered at the origin with radii $t_i\in (0,\sqrt{n-1})$,
  where $\{t_i\}$ is an increasing sequence converging to
  $\sqrt{n-1}$.
  These polytopes are characterized by $n^2$
  inequalities in at most $n^2$ variables.
  The existence of fpras for
  computing volumes of convex bodies gives rise to a semi-fpras of order
  $O^*(n^{14})$ at most to find
  if given two undirected graphs are isomorphic.
  \\



 \end{abstract}


\section{Introduction}

 Let $G_1=(V,E_1),G_2=(V,E_2)$ be two simple undirected graphs,
 where $V$ is the set of vertices of cardinality $n$ and
 $E_1,E_2\subset V\times V$ the set of edges. $G_1$ and $G_2$
 are called \emph{isomorphic} if there exists a bijection $\sigma:V\to V$
 which induces the corresponding bijection $\tilde
 \sigma:E_1\to  E_2$.
 The graph isomorphism problem, abbreviated here as \emph{GIP},
 is the computational complexity of
 determination if $G_1$ and $G_2$ are isomorphic.
 Clearly the \emph{GIP} in the class \emph{NP}.
 It is one of a very small number of problems whose complexity
 is unknown \cite{GJ, KST}.  For certain graphs it was known that the complexity
 of \emph{GIP} is polynomial \cite{BGM,Bod,FM,HW,Luk,Mil}.

 The graph isomorphism problem is a special case of
 permutational similarity of two $n\times n$ real values matrices
 $A,B\in\R^{n\times n}$, or more generally $n\times n$ matrices
 with entries in any ring with identity.  Namely, let
 $\cP_n\subset \R^{n\times n}$ be the group of permutation
 matrices.  Does there exists $P\in\cP_n$ such that
 $B=PAP\trans$?

 Using the notion of coherent algebras, as a tool to identify
 nontrivial pairs of matrices $A,B\in\R^{n\times n}$ which may
 be permutationally similar \cite{Fri89}, we first show
 that a nontrivial permutational similarity of
 $A,B\in\R^{n\times n}$ can be polynomially reduced to an isomorphism
 problem of two regular undirected connected multi-graphs with the same degree,
 (self-loops allowed),
 and the same characteristic polynomial.
 Assume that these two graphs represented by symmetric
 $A,B\in \rS_n(\Z_+)$ its rows sum equal to $N$.

 Denote by $\1=(1,\ldots,1)\trans$, and let $\rM_n\subset
 \R^{n\times n}$ be the space of matrices with zero row and column
 sum:
 \begin{equation}\label{defMn}
 X\1=X\trans \1=\0, \quad X=[x_{ij}]\in\R^{n \times n}.
 \end{equation}
 Note that $\rM_n$ is $(n-1)^2$ dimensional subspace of
 $\R^{n\times n}$.  Denote by $\Omega_n\subset \R^{n \times n}$
 the convex set of doubly stochastic matrices.  Note that
 $Y\in\Omega_n$ if and only if $Y=X+J_n, J_n:=\frac{1}{n}\1\1\trans$,
 where $X\in \rM_n$ and each entry of $X$ satisfies $t\ge -\frac{1}{n}$.
 It is easy to see that $\|X\|_F:=\sqrt{\tr X X\trans}\le
 \sqrt{n-1}$ if $-J_n\le X\in\rM_n$, and equality holds if and only
 if $X=P-J_n$ for some $P\in\cP_n$.

 For any $S,T\in \R^{n\times n}$ we define the following
 subspace of matrices and a corresponding
 bounded polytope:
 \begin{equation}\label{defpolST}
 \bP_0(S,T):=\{X\in\rM_n,\; SX-XT=0\},\;
 \bP(S,T):=\{X\in\bP_0(S,T),\; X\ge -J_n\}.
 \end{equation}
 Denote by $\|\bP(S,T)\|_F:=\max_{X\in\bP(S,T)} \|X\|_F$, the
 radius of $\bP(S,T)$.
 Thus two regular multi undirected graphs $G,H$, with the the same
 number of vertices and edges, are isomorphic
 if and only if $\|\bP(A,B)\|_F=\sqrt{n-1}$, where $A,B$ are
 the representation matrices of $G,H$ respectively.
 However, it is known that finding the radius of a convex set
 is $NP$-hard \cite{LS93}.  (Note that the diameter of a balanced
 convex set $K$, i.e. $-K=K$, is twice its radius.)

 The main result of this paper is
 \begin{theo}\label{maintheorem}
 Let $G,H$ be two regular multi undirected graphs, with the the same
 number of vertices $n$ and edges $e$.  Denote by $A,B\in \Z_+^{n\times n}$ the
 representation matrices of $G,H$ respectively.  The the
 following statements are equivalent.
 \begin{enumerate}
 \item \label{maintheoremcon1} $G$ and $H$ are isomorphic.
 \item\label{maintheoremcon2}
 The dimension of the convex sets $\bP(A,A),\bP(A,B),\bP(B,B)$ are
 equal.
 Furthermore, for each $t\in (0,\sqrt{n-1}]$ the volumes of the intersection
 of the above three polytopes
 with the ball of radius $t$
 centered at $0$ are the equal.

 \item\label{maintheoremcon3}
 The dimension of the convex sets $\bP(A,A),\bP(A,B),\bP(B,B)$ are
 equal.
 Furthermore,
 for a given sequence of balls $t_i\in (0,\sqrt{n-1})$,
 where $\{t_i\}$ is an increasing sequence converging to $\sqrt{n-1}$,
 the volumes of the intersection of the above three polytopes
 with each ball of radius $t_i$
 centered at $0$ are the equal.

 \end{enumerate}

 \end{theo}

 The main argument of the proof of this is theorem follows straightforward
 from the observation that $I_n-J_n\in \bP(A,A)$, i.e. $\|\bP(A,A)\|_F=\sqrt{n-1}$.

 We show that for the convex sets $\bP(A,A), \bP(A,B),\bP(B,B)$ one can apply the
 known results, which give fully randomized polynomial
 approximation scheme for computing the volumes of the
 intersection of these set with a ball of radius $t\in
 (\frac{1}{n},\sqrt{n-1})$, e.g \cite{DFK91, LS93, KLS97, LV06}.
 Combining these results we obtain
 some algorithms for testing the volume conditions given by
 Theorem \ref{maintheorem}.  Recall that the problem of
 finding the exact volume of a polytope in $\R^m$, given by a polynomial
 number of affine inequalities in $m$, is $\#P$-hard
 \cite{DF88}.  We hope that this approach will lead to a fpras to
 determine if given two graphs are isomorphic.

 We now summarize briefly the contents of this paper.
 In \S2 we discuss the notion of coherent algebras and their
 relations to the graph isomorphism problem.  In \S3 we
 construct the polytopes $\bP(A,A),\bP(A,B),\bP(B,B)$ which are
 intimately related to a permutational similarity of $A,B$,
 assumed to be scaled doubly stochastic with the same
 row sums.  We give an outline of the proof of Theorem
 \ref{maintheorem}.  In \S4 we outline a semi-fpras to find if
 given $A,B$ are permutationally similar, which is based on the
 fpras for computing volume of convex sets.

 \section{Coherent algebras}

 A subalgebra $\cC\subset \R^{n\times n}$ is called a
 \emph{coherent algebra}, if it is closed under
 the transposition and entry-wise multiplication of two
 matrices, and
 contains $I_n=[\delta_{ij}],
 J_n=[\frac{1}{n}]$, the identity matrix and the doubly
 stochastic matrix with equal entries.
 (Denote by $A\circ B=[a_{ij}]\circ [b_{ij}]$ the entry-wise
 product $[a_{ij}b_{ij}]$.)  We now briefly survey the main
 properties of coherent algebra used her.  Our main source
 is our paper \cite{Fri89}.  Additional references for the properties
 of coherent algebras cited explicitly, where needed.

 A trivial coherent algebra is an algebra of
 dimension $2$ spanned by $I,J$.  Coherent algebras of
 dimension $3$ are induced either by strongly regular graphs, or
 by Hadamard matrices.
 A coherent algebra $\cC$ has a canonical basis consisting of
 $(0,1)$ matrices $E_1,\ldots, E_d$. Each $E_i$ is either symmetric or
 asymmetric, i.e. $E_i\circ E_i\trans=0$, and balanced,
 i.e. the nonzero rows and columns of $E_i$ are equal to $r_i$
 and $c_i$ respectively.  Furthermore,
 $E_i\circ E_j=0$ for $i\ne j$, and $\sum_{i=1}^d E_i=\1\1\trans$.  $\cC$ is characterized by the
 tensor $T(\cC)=[t_{i,j,k}]\in\Z_+^{d\times d\times d}$.
 \begin{equation}\label{deftenca}
 E_iE_j=\sum_{k=1}^d t_{i,j,k}E_k.
 \end{equation}

 Any $A\in\R^{n\times n}$ induces the minimal coherent subalgebra
 $\cC(A)\subset \R^{n\times n}$ containing $A$.
 One finds in polynomial time the canonical basis $E_1,\ldots,E_d\in
 \{0,1\}^{n\times n}$ of
 $\cC(A)$.  (\cite[Lemma 3.1]{Fri89} yields that one needs at
 most $17n^{10}$ flops.)
 If $B\in\R^{n\times n}$ is permutationally similar
 to $A$, it follows that $\cC(B)$ is strongly isomorphic to
 $\cC(A)$.  So $\cC(B)$ has the canonical basis
 $F_1,\ldots,F_d$, such that $F_i=PE_iP\trans, i=1,\ldots,d$
 for a corresponding $P\in\cP$.  These equalities induces the strong isomorphism
 $\iota:\cC(A)\to\cC(B)$ given by $\iota(E_i)=F_i,
 i=1,\ldots,d$.

 Thus for $A,B\in\R^{n\times n}$ to be permutationally similar
 we must have an isomorphism $\iota:\cC(A)\to\cC(B)$, such that
 $\iota(E_i)=F_i, i=1,\ldots,n$.  ($\iota$ is an isomorphism of
 two algebras, which preserve the transposition and entry-wise
 multiplication, i.e. $\iota(U\trans)=\iota(U)\trans, \iota
 (U\circ V)=\iota(U)\circ\iota(V)$.)
 In particular, $T(\cC(A))=T(\cC(B))$, and this condition is
 essentially equivalent to isomorphism of $\cC(A)$ and $\cC(B)$.

 The existence or nonexistence
 of such isomorphism is determined in a polynomial time in
 $O(n^{10})$.
 It is possible that the isomorphism of coherent
 algebras does not imply the strong isomorphism.
 \begin{theo}\label{equivconpsim}
 Let $A,B\in\R^{n\times n}$.  Assume that the coherent algebras
 $\cC(A)$ and $\cC(B)$ are isomorphic, i.e. $\iota:\cC(A)\to
 \c(B)$ is an isomorphism of coherent algebras.
 Then there exists $A_1\in\cC(A), B_1\in\cC(B)$ with the
 following properties: $\iota(A_1)=\iota(B_1)$; $A_1$ and
 $B_1$ are symmetric matrices with positive integer entries
 whose values are less than $n^3$;
 each row sum of $A_1$ and $B_1$ is equal to $N$;
 $A_1$ and $B_1$ have the same characteristic polynomial.
 Furthermore, $A$ and $B$ are permutationally similar if and
 only if $A_1$ and $B_1$ are permutationally similar.
 \end{theo}
 \outproof  Assume $E_1,\ldots, E_h, h\ge 1$ are all the
 diagonal matrices in the canonical basis $E_1,\ldots,E_d$.
 Consider $A_2:=\sum_{i=h+1}^{d} m_i E_i$, where
 $m_2,\ldots,m_d\in\N$ satisfy the condition $m_i\ne m_j$ unless
 $E_i\trans =E_j$.  (If  $E_i\trans =E_j$ we let $m_i=m_j$.)
 The number of distinct integers in $\{m_{h+1},\ldots,
 m_d\}$ is $p\le \frac{n(n-1)}{2}$.  Hence we can assume that
 the set of distinct integers in $\{m_{h+1},\ldots, m_d\}$ is
 $\{1,\ldots,p\}$.  Let $N-1$ be the maximal row sum of $A_2$.
 Then $A_1=A+D$, where $D$ is the diagonal matrix such that
 each row of $A_2$ equals to $N$.  The results of \cite{Fri89}
 imply that $A_1\in \cC(A)$.  Set $B_1=\iota(A_1)$.  Then all
 other claims of the theorem follow straightforward  from the
 results in \cite{Fri89}.  \qed

 Note that $A_1,B_1$ are representation matrices of two regular
 undirected multi-graphs $G,H$ with self loops, with the same numbers of vertices,
 edges, and the same characteristic polynomials.  Furthermore,
 $G$ and $H$ are connected.
 In the rest of the paper we assume that $A=A_1, B=B_1$.

 It is possible to show, that by increasing the entries of $A_1$,
 which are still are of
 order $O(n^K)$, that in addition to the above conditions on $A_1$,
 $A_1$ is generic in $\cC(A)$.
 That is, the multiplicity of each eigenvalue of
 $A_1$ is the minimal possible for any symmetric matrix
 $S\in \cC(A)$.

 More generally, any set of $A_1,\ldots,A_k\in\R^{n\times n}$
 induces a minimal coherent subalgebra
 $\cC(A_1,\ldots,A_k)\subset\R^{n\times n}$ which contains these matrices.
 It is obtained
 by the following process.  First, express each matrix $A_i$ as
 a unique linear combination of $(0,1)$ matrices with pairwise
 distinct coordinates. This gives rise to $T_1,\ldots, T_k\in
 \{0,1\}^{n\times n}$.  (At the first step $T_1=I_n,
 T_2=\1\1\trans$.)  By considering  the subspace spanned by
 all nonzero $(0,1)$ matrices
 of the form $T_i\circ T_j, T_i\trans\circ T_j, T_i\trans\circ
 T_j\trans, i,j=1,\ldots,k$ we obtain a subspace $\U_1\subset \R^{n\times n}$
 spanned by
 $(0,1)$ matrices $R_1,\ldots,R_l$ with disjoint support such that their sum is
 equal to $\1\1\trans$.  We now consider the set of matrices
 $R_iR_j, i,j=1,\ldots,l$, whose span $\U_2$ includes
 $R_1,\ldots,R_l$.  Apply the previous algorithm to these $l^2$
 matrices to obtain a $(0,1)$ basis in $\U_2$ which is a refinement
 of the basis $R_1,\ldots, R_l$.  After  $p\le n^2$ steps we
 will have that $\U_{p}=\U_{p+1}$.  Then $\cC(A_1,\ldots,A_k)=\U_p$.

 We say that the set  $A_1,\ldots,A_k\in \R^{n\times n}$ is
 permutationally similar to $B_1,\ldots,B_k\in \R^{n\times n}$
 if $B_i=PA_iP\trans$ for $i=1,\ldots,k$ and some $P\in\cP_n$.
 As in the case $k=1$ the nontrivial permutational similarity
 induces an isomorphism of the coherent algebras
 $\iota:\cC(A_1,\ldots,A_k)\to\cC(B_1,\ldots,B_k)$, such that
 $\iota(A_i)=B_i,i=1,\ldots,k$.
 As in the case $k=1$, the problem of nontrivial permutation
 similarity of $A_1,\ldots,A_k$ and $B_1,\ldots,B_k$ can be
 reduced to permutational similarity of two symmetric scaled
 doubly stochastic matrices $A,B\in \N^{n\times n}$ with the
 same characteristic polynomial.

 \section{Convex polytopes associate with GIP}

 Let $\Omega_n\subset \R_+^{n\times n}$ be the convex set of
 $n\times n$ doubly stochastic matrices.
 Recall that $\Omega_n=\{X\in\rM_n,\;X\ge -J_n\}$.
 We say that $A\in\R_+^{n\times n}$ is a scaled doubly
 stochastic matrix if $A=aC$ for some $a>0$ and $C$ doubly
 stochastic.
 \begin{lemma}\label{dimPAA}
 Let $A\in\R_+^{n\times n}, n\ge 2$ be an irreducible symmetric scaled
 doubly stochastic matrix.  Assume that $A$ has $\mu+1\ge 2$
 distinct eigenvalues
 $\lambda_0>\lambda_1>\ldots>\lambda_{\mu}$, where $m_i$ is the
 multiplicity of $\lambda_i$ for $i=0,\ldots,\mu$.  ($m_0=1$.)
 Then the dimension of the subspace $\bP_0(A,A)$ and
 the polytope $\bP(A,A)$, given by
 (\ref{defpolST}), is $\Delta:=\sum_{i=1}^{\mu} m_i^2$.

 Assume that $B\in\R_+^{n\times n}$ is an irreducible symmetric scaled
 doubly stochastic matrix having the same row sums as $A$.
 Then the following are equivalent
 \begin{enumerate}
 \item $A$ and $B$ have the same characteristic polynomial.
 \item
 The three subspaces $\bP_0(A,A),\bP_0(A,B),\bP_0(B,B)$
 have the same dimension.
 \end{enumerate}
 \end{lemma}
 \proof  Since $A$ is irreducible, its Perron-Frobenius root,
 $\lambda_0$ is the largest eigenvalue of multiplicity $1$.
 As $A$ is scaled doubly stochastic, $A=\lambda_0 C$ for some
 symmetric stochastic matrix $C$.  So $A\1=\lambda_0\1$.
 Choose an orthonormal basis of $\R^n$ consisting of
 orthonormal eigenvectors $\x_0=\frac{1}{\sqrt{n}}\1,\x_1,\ldots,\x_{n-1}$,
 corresponding to the eigenvalues
 $\lambda_0>\lambda_1>\ldots>\lambda_{\mu}$.  Let
 $Q\in\R^{n\times n}$ be the orthogonal matrix whose columns
 are $\x_0,\ldots,\x_{n-1}$.  Then $Q\trans A
 Q$ is the block diagonal matrix $\oplus_{i=0}^{\mu} \lambda_i I_{m_i}$.
 Observe that
 \begin{equation}\label{QfromMn}
 Q\trans \rM_nQ=\{0_{1\times 1} \oplus Z,\;Z\in \R^{(n-1)\times
 (n-1)}\}.
 \end{equation}
 Recall next that any commuting matrix with $Q\trans AQ$ has the block
 diagonal form $\oplus_{i=0}^{\mu} U_i$ where $U_i\in
 \R^{m_i\times m_i}$ for $i=0,\ldots,\mu$.  Hence
 \begin{equation}\label{QformP0AA}
 Q\trans \bP_0(A,A)Q=\{0_{1\times 1}\oplus_{i=1}^{\mu} U_i,\;
 U_i\in\R^{m_i\times m_i}, i=1,\ldots,\mu\}.
 \end{equation}
 Thus $\dim \bP_0(A,A)=\Delta$.  Since $0_{n\times n}\in
 \bP_0(A,A)$ is an interior point of $\bP(A,A)$ it follows that
 $\dim\bP(A,A)=\Delta$.

 Assume first that $B$ has the same characteristic polynomial as $A$.
 Since $A,B$ are symmetric, there exists an orthogonal matrix
 $Q_1$, with he columns $\x_0,\y_1,\ldots,\y_{n-1}$, such that
 $Q_1\trans BQ_1=\oplus_{i=0}^{\mu} \lambda_i I_{m_i}$ where t.  Hence
 \begin{equation}\label{eqPAAPAB}
 \bP_0(A,B)= \bP_0(A,A)Q_2\trans, \quad Q_2=Q_1Q\trans,\;B=Q_2 A Q_2\trans.
 \end{equation}
 So $\dim\bP_0(A,B)=\dim \bP_0(B,B)=\Delta$.

 Assume that $\dim\bP_0(A,A)=\dim\bP_0(A,B)=\dim \bP_0(B,B)$.
 Hence the dimension of the following three subspaces in $\R^{n\times
 n}$:
 $\{X,\;AX-XA=0\},\;\{X,\;AX-XB=0\},\;\{X,\;BX-XB=0\}$
 are equal.  Therefore $A$ and $B$ are similar \cite{Fri80}.
 \qed\\
 \textbf{Proof of Theorem }\ref{maintheorem}.
 Observe first that if $U,V\in\R^{n\times n}$ are orthogonal
 matrices, then the transformation $X\mapsto UXV$ is an
 orthogonal transformation on $\R^{n\times n}$.
 In particular, the ball $\rB(0,t)\subset \R^{n\times n}$,
 centered at $0$ of radius $t$ satisfies the equality
 $U\rB(0,t)V=\rB(0,t)$ for any orthogonal $U,V$.

 Assume first that $A,B$ are similar, i.e $B=Q_2 A Q_2\trans$ for some orthogonal
 $Q_2$.  Then for any $t>0$
 \begin{eqnarray*}
 \rB(0,t)\cap\bP_0(A,B)=(\rB(0,t)\cap\bP_0(A,A))Q_2\trans,\\
 \rB(0,t)\cap\bP_0(B,B)=Q_2(\rB(0,t)\cap\bP_0(A,A))Q_2\trans.
 \end{eqnarray*}
 Suppose furthermore that $A$ and $B$ are permutationally
 permutationally similar, i.e $Q_2=P\in\cP_n$.  (So \ref{maintheoremcon2}
 holds.)  Then (\ref{eqPAAPAB}) yields that
 $\bP(A,B)=\bP(A,A) P\trans, \bP(B,B)=P\bP(A,A)P\trans$.
 In particular
 \begin{eqnarray*}
 \rB(0,t)\cap\bP(A,B)=(\rB(0,t)\cap\bP(A,A))P\trans,\\
 \rB(0,t)\cap\bP(B,B)=P(\rB(0,t)\cap\bP(A,A))P\trans.
 \end{eqnarray*}
 Hence, above intersections have the same volume for any $t>0$.
 This proves the conditions
 \ref{maintheoremcon2}-\ref{maintheoremcon3}.

 Recall that $\|\bP(A,B)\|_F\le \sqrt{n-1}$, and equality holds
 if and only if $(\cP_n-J_n)\cap \bP(A,B)\ne \emptyset$, i.e. $A$ and
 $B$ are permutationally similar.
 Observe next that since $I_n-J_n\in \bP(A,A)\cap \bP(B,B)$ it
 follows that $\|\bP(A,A)\|_F=|\bP(B,B)\|_F =\sqrt{n-1}.$
 Hence, the volumes of $\rB(0,t)\cap\bP(A,A),
 \rB(0,t)\cap\bP(B,B)$ increase in the interval
 $(0,\sqrt{n-1})$.

 Assume that \ref{maintheoremcon3} holds.  Since the volumes of
 $\rB(0,t_i)\cap\bP(A,A), i=1,\ldots,$ form an increasing,
 it follows that the volumes of $\rB(0,t_i)\cap\bP(A,B), i=1,\ldots,$
 form an increasing sequence.  Hence $\|\bP(A,B)\|=\sqrt{n-1}$.
 So $A$ and $B$ are permutationally similar.
 Clearly, the same arguments imply that the condition \ref{maintheoremcon2}
 implies the permutational similarity of $A$ and $B$.  \qed

 \section{A semi-fpras for graph isomorphism}

 We now point out how to apply the existing fully polynomial randomized
 approximation schemes to compute a volume of a
 convex sets, e.g. \cite{KLS97}.
 To do that it would be convenient to map the three convex
 polytopes $\bP(A,A)$, $\bP(A,B)$, $\bP(B,B)$ of dimension $\Delta$,
 to one ambient space $\R^{\Delta}$ by
 by three different linear transformations:
 \begin{equation}\label{3lintran}
 T_1:\bP_0(A,A)\to\R^{\Delta},\;
 T_2:\bP_0(A,B)\to\R^{\Delta},\;
 T_3:\bP_0(A,A)\to\R^{\Delta},
 \end{equation}
 such that each $T_i$ preserves the inner product.
 We demonstrate for $T_2$. Choose an orthonormal basis
 $W_1,\ldots,W_{\Delta}$ in $\bP_0(A,B)$.  Then
 $T(W_i)=(\delta_{1i}, \ldots,\delta_{n\Delta})\trans,
 i=1,\ldots,\Delta$.  It is straightforward to show that
 $T_1(\bP(A,A)), T_2(\bP(A,B)), T_3(\bP(B,B))$ are polytopes
 $\X_1,\X_2,\X_3\subset \R^{\Delta\times \Delta}$, which are
 given as follows.  There exists nonzero vectors $\u_{(1,1),l},\ldots,\u_{(n,n),l}\in
 \R^{\Delta}, l=1,2,3$ such that
 \begin{equation}\label{defXi}
 \X_l=\{\x\in \R^{\Delta}, \;\u_{(i,j),l}\trans \x\ge -\frac{1}{n},
 i,j=1,\ldots,n\}, \quad l=1,2,3.
 \end{equation}
 One can compute the vectors $\u_{(i,j),l}$ in polynomial time in
 $n$.  Note that the inequality
 $x_{ij}\ge -\frac{1}{n}$ is equivalent to $\u_{(i,j),l}\trans \x\ge
 -\frac{1}{n}$ in the orthonormal basis of the corresponding
 linear space $\bP_0(A,A),\bP_0(A,B),\bP_0(B,B)$.

 As we explain in the next section the permutational similarity
 of $A$ and $B$ is equivalent to the existence of an
 orthonormal matrix $O\in \R^{\Delta\times \Delta}$ such that
 $O\{\u_{(1,1),1},\ldots,\u_{(n,n),1}\}=\{\u_{(1,1),2},\ldots,\u_{(n,n),2}\}$.
 (Similar statement holds for

 \noindent
 $\{\u_{(1,1),3},\ldots,\u_{(n,n),3}\}$,
 $\{\u_{(1,1),2},\ldots,\u_{(n,n),2}\}$.)

 It is trivial to see that the polytopes
 $\bP(A,A),\bP(A,B),\bP(B,B)$ contain the ball of radius
 $\frac{1}{n}$ centered at the origin.  Theorem
 \ref{maintheorem} yields that $A$ and $B$ are permutationally
 similar if
 \begin{equation}\label{eqvolX123}
 \vol(\rB(0,t)\X_1)=\vol(\rB(0,t)\X_2)=\vol(\rB(0,t)\X_3)
 \textrm{ for each } t\in (\frac{1}{n}, \sqrt{n-1}).
 \end{equation}

 We now suggest a probabilistic test of the
 above equalities, for a finite number of values $t\in (\frac{1}{n},
 \sqrt{n-1}$ with a relative error $\varepsilon$ and with
 probability $1-\eta$.  In the random algorithms suggested in
 \cite{DFK91,LS93, KLS97}, adopted to find the volumes of
 $\X_i,i=1,2,3$,
 one considers the intersection of the sequence of balls of
 radii:
 \begin{equation}\label{seqradtj}
 t_j=\frac{2^{\frac{j}{N}}}{n}, \quad j=0,\ldots, M=\lceil
 N\log_2 n\sqrt{n-1}\rceil.
 \end{equation}
 Here $N$ can be chosen as $\Delta$, as  in \cite{DFK91,LS93,
 KLS97}, or if we want more points we can take $N=n^c \Delta$
 for some $c>0$.
 Let $\X_{j,i}:=\rB(t_j)\cap \X_i, j=0,\ldots,M, i=1,2,3$.
 For each $\X_{j,i}$ one generates $p=400\varepsilon^{-2}N\log
 N$ random points from certain distribution, e.g.
 \cite[\S6]{KLS97}.  Then
 $\frac{\vol(\X_{j,i})}{\vol(\X_{j-1,i})}$ is estimated by the
 fraction of number of the sampled points in $\X_{j,i}$ to the
 number of the sampled in $\X_{j,i}$ which are in $\X_{j-1,i}$.
 By Theorem \ref{maintheorem}, if $A$ and $B$ are permutationally
 similar, we must have the equalities
 \begin{equation}\label{voleqgrphiso}
 \frac{\vol(\X_{j,1})}{\vol(\X_{j-1,1})}=\frac{\vol(\X_{j,2})}{\vol(\X_{j-1,2})}
 =\frac{\vol(\X_{j,3})}{\vol(\X_{j-1,3})},\;
 \vol(\X_{j,1})=\vol(\X_{j,2})=\vol(\X_{j,3})
 \end{equation}
 for $j=1,\ldots,M$.
 Hence, in our process of estimating the volumes of
 $\X_1,\X_2,\X_3$ we test the above equalities within relative
 error $\varepsilon$.  If all the above equalities hold within
 the relative error $\varepsilon$, we declare that the matrices
 $A,B$ are $\epsilon,\eta$ permutationally similar.  If one of
 the equalities in (\ref{voleqgrphiso}) fails with respect to relative
 error $\varepsilon$, we have two options.  Either declare that the matrices
 $A,B$ are not $\varepsilon,\eta$ permutationally similar, or
 retest this equality with a smaller $\varepsilon$ and $\eta$.
 If all the retested equalities hold, then we declare that
 $A,B$ are $\epsilon,\eta$ permutationally similar.
 Otherwise, we declare that the matrices
 $A,B$ are not $\varepsilon,\eta$ permutationally similar.

 Note that each oracle query if a point $\x \in \rB(t)\cap
 \X_i$ needs $\Delta n^2\le n^4$ flops, since the dot
 product in $\R^{\Delta}$ need $\Delta$ flops.  Since the
 randomized algorithm suggested in \cite{KLS97} is of order
 $O^*(\Delta^5)$ we see that the our algorithm for checking
 the graph isomorphism, or permutational similarity of $A,B\in
 \R^{n\times n}$, is of order $O^*(n^{14})$ at most.

\end{document}